\documentclass[aps,showpacs,pra,preprint]{revtex4}

\usepackage{latexsym}
\usepackage{bbm}
\usepackage{amssymb}
\usepackage{graphicx}
\usepackage{graphics}

\newcommand{\beq}{\begin{equation}}
\newcommand{\eeq}{\end{equation}}
\newcommand{\bea}{\begin{eqnarray}}
\newcommand{\eea}{\end{eqnarray}}
\newcommand{\bra}[1]{\left\langle #1 \right\vert}
\newcommand{\ket}[1]{\left\vert #1 \right\rangle}
\newcommand{\al}{\alpha}

\newcommand{\R}{\hat{R}}
\newcommand{\F}{\cal F}
\newcommand{\kr}{\hat{E}}
\newcommand{\den}{\hat{\rho}}

\begin{document}

\title{Quantum error correction may delay, but also cause, entanglement sudden death}
\author{Isabel Sainz}
\author{Gunnar Bj\"ork}
\affiliation{School of Information and Communication Technology,
Royal Institute of Technology (KTH), Electrum 229, SE-164 40 Kista,
Sweden}

\date{\today}

\begin{abstract}
Dissipation may cause two initially entangled qubits to evolve into
a separable state in a finite time. This behavior is called
entanglement sudden death (ESD). We study to what extent quantum
error correction can combat ESD. We find that in some cases quantum
error correction can delay entanglement sudden death but in other
cases quantum error correction may cause ESD for states that
otherwise do not suffer from it. Our analysis also shows that
fidelity may not be the best measure to compare the efficiency of
different error correction codes since the fidelity is not directly
coupled to a state's remaining entanglement.
\end{abstract}

\pacs{03.67.Mn, 03.65.Yz, 03.67.Pp}

\maketitle

\section{Introduction}

The view of entanglement has gone from being a consequence of
quantum theory that ``no reasonable definition of reality could be
expected to permit'' \cite{EPR} to a physical resource on which
quantum technologies are based \cite{nielsen}. For that reason it is
important not only to understand entanglement dynamics, but it is
desirable to be able to manipulate entanglement states in a
practical way.

Entanglement dynamics has been studied in several contexts. When
interacting with a reservoir, one would naively expect the
entanglement between two systems to vanish asymptotically. However,
inspired by the entanglement breaking quantum channel theorems
\cite{breaking}, Di\'osi showed that if two systems are initially
entangled, and they are subjected independently (or only one of
them) to a depolarizing channel, the systems will become
disentangled in a finite time \cite{diosi}. This result was
subsequently extended for continuous variables in \cite{dodd}.
Around the same time, Yu and Eberly \cite{yu1} found that two
initially entangled qubits placed in their own cavities can become
disentangled via spontaneous emission at a rate larger than the
decoherence rate. For certain initial states, the time of
disentanglement is finite, a phenomenon termed ``entanglement sudden
death'' (ESD) \cite{esd}. Lately, it has been found that ESD is
present in different scenarios for certain classes of states, (see,
for example, \cite{esd1,xstates,esd2,eberly} and the references
therein). Moreover, this phenomenon was recently demonstrated in an
experiment \cite{almeida}.

A worry, expressed in the literature, is that when an entangled
state has suffered from entanglement sudden death due, e.g., to
dissipation, and become disentangled, no (local) recovery operation
can restore the state \cite{eberly}. This means that the initial
quantum resource carried by the state is irrevocably lost. ESD has
hence been seen as a potential hurdle, or even impasse, for the
development of, e.g., large scale quantum computing.

To combat the decoherence of quantum states, several schemes have
been proposed during the last decade. When the considered subsystems
are coupled to a collective environment, one can make use of
decoherence-free subspaces \cite{dfs}. Another technique, that does
not necessarily require that the subsystems are coupled to the same
environment, is the open-loop control \cite{bangbang}. However,
subsequently it was shown that the latter technique is the dynamical
restoration equivalence of the decoherence-free subspaces
\cite{dfsrestore,qecdfs}.

Another option to deal with noisy or dissipative channels is quantum
error correction (QEC). In this scheme one can restore a
``distorted'' state by the introduction of redundancy by coding
\cite{shor,steane,paz,5qubit,QECconditions,qec,plenio97,leung,FletcherN,Fletcher}.
In general, QEC works in three stages: At the first stage, every
physical qubit is encoded onto a logical qubit (code) composed of
several physical qubits. The logical qubits then interact with a
non-ideal (noisy) channel, that ``distorts'' the state. This
distortion is unwanted, but will occur in most realistic channels.
At the second stage, one performs a measurement to find the error
syndrome, i.e., what kind of error the channel caused and in which
physical qubit it occurred. In the third and final stage one
restores the distorted logical state into the initial logical state
by applying a recovery operation, conditioned on the information
given by the error syndrome. Several well known error correcting
codes exist \cite{shor,steane,paz,5qubit}, and these enables one to
correct a ``general'' error (e.g., represented by the Pauli
operators) in a single qubit. (If two or more qubits are erroneous,
more sophisticated codes must be used to restore the state.)
Typically the loss of fidelity after correction is of the order of
$\gamma^2$, where $\gamma$ is the probability of having one error in
one of the physical qubits. This result should be compared with the
fidelity without coding and error correction, where the loss of
fidelity scales as $\gamma$.

In open quantum systems not all the possible errors can be
represented by the discrete Pauli operators (that represent
bitflips, phaseflips and both), e.g., amplitude damping. The codes
like \cite{shor,steane,paz,5qubit} are not designed to correct
open-system errors. Such errors do not fulfil the so-called QEC
criteria \cite{5qubit,QECconditions}, and they can therefore not be
exactly corrected. To address this problem, a channel-adapted
quantum error correction scheme has been proposed in
\cite{plenio97}, where a code is presented that is able to correct a
single general (discrete Pauli) error and is at the same time
perfectly adapted to correct errors due to spontaneous emission.
Concurrently, it was shown that small deviations from the QEC
criteria will not change the fidelity qualitatively (in terms of the
order of $\gamma$), and sufficient conditions for an approximate QEC
were given \cite{leung}. Recently, numerical algorithms to generate
channel-adapted recovery operations using well known codes such as
\cite{shor,steane,paz,5qubit} were presented \cite{FletcherN}. In
\cite{Fletcher} channel-adapted codes for an amplitude damping
channel are developed, where the encoding is described in the
stabilizer formalism. Different recovery operations are compared,
specially optimal recovery and channel-adapted stabilizer recovery.
The conclusion from \cite{Fletcher} is that the improvement in
fidelity by the optimal recovery is small compared to the cost of
implementation.

In this paper, we will investigate to what extent ESD can be
combatted using quantum coding adapted for dissipative channels. The
hope is that the coding, which spreads the initial two-qubit
entanglement over many qubits, may prevent ESD altogether. Somewhat
surprisingly, the result is that while coding and error correction
will indeed help for small amounts of dissipation, quantum coding
actually makes matter worse when the dissipation is large. In
particular, states that do not suffer from ESD when uncoded will, in
coded form, display ESD.

\section{The model}

Entanglement sudden death is cased by dissipation. Consequently we
will consider an amplitude damping channel, that for one qubit can
be represented in terms of the Kraus operators :
\begin{displaymath}
\kr_0=\left(
\begin{array}{cc}
1&0\\
0&\sqrt{1-\gamma}
\end{array}\right),\qquad
\kr_1=\left(
\begin{array}{cc}
0&\sqrt{\gamma}\\
0&0
\end{array}\right),
\end{displaymath}
where $\gamma$ is the damping, $\kr_0$ is the one-qubit no-jump
operator that leaves the ground state $\ket{0}$ unchanged, but
reduces the probability amplitude of the excited state $\ket{1} $
with the factor $1-\gamma$, and $\kr_1$ represents the jump operator
that transforms the state $\ket{1}$ into the state $\ket{0}$ with
probability $\gamma$.

To investigate to what extent a two qubit state can be ``protected''
against amplitude damping and in particular to ESD, we will use the
$[4,1]$ code introduced in \cite{leung}, where the notation $[4,1]$
denotes coding of each initial qubit onto a logical qubit consisting
of four physical qubits. (We have also tested $[5,1]$ and $[6,2]$
codes with similar results.) The $[4,1]$ code encodes an arbitrary
(single qubit) state $\ket{\varphi}=\cos\al\ket{1}+\sin\al\ket{0}$
into the logical state
$\ket{\varphi}_L=\cos\al\ket{1}_L+\sin\al\ket{0}_L$, were the
codewords $\ket{0}_L$ and $\ket{1}_L$, are given by the four qubit
code \cite{leung,Fletcher}, \bea
\ket{0}_L&=&\left(\ket{0000}+\ket{1111}\right)/\sqrt{2},\label{ceroL}\\
\ket{1}_L&=&\left(\ket{0011}+\ket{1100}\right)/\sqrt{2}\label{unoL},\eea
where here, and below, we will use the notation
$\ket{0000}\equiv\ket{0}\otimes\ket{0}\otimes\ket{0}\otimes\ket{0}$.

For simplicity we will suppose that all the physical qubits have the
same damping probability $\gamma$. We will also assume that the
amplitude damping channels are independent, which means that if the
amplitude damping channel $\kr$ acts on the one-qubit density matrix
$\den$ as \beq \kr(\den)=\kr_0\den \kr_0^{\dag}+\kr_1\den
\kr_1^{\dag},\eeq the many-qubit amplitude damping channel will be
described by the tensorial products of the corresponding Kraus
operators, e.g., for two qubits, the channel will be described by
operators of the form \beq\kr_{mn}=\kr_m\otimes
\kr_n\label{multipledamp}\eeq where $m,n=0,1$, and the first(second)
operator acts on the first(second) qubit.

In our assumed channel we have a conditional evolution, because the
ground state is not affected by the damping whereas the excited
state is. Unfortunately, some well-know QEC protocols such as those
in \cite{shor,steane,paz,5qubit} are not intended for this kind of
channels. Nevertheless, using the same codewords one can find
suitable syndrome measurements and recovery operators that do not
fulfil the QEC conditions \cite{QECconditions}, but the
``approximate'' ones \cite{leung}. Hence, the $[4,1]$ code in
(\ref{ceroL}) and (\ref{unoL}) can only ``approximately'' correct no
jump and one qubit jumps, in the sense that the distortion caused by
the conditional evolution is of the second order in the damping
parameter $\gamma$ \cite{leung}.

As was pointed out in \cite{Fletcher}, the states reached from
$\ket{0}_L$ and $\ket{1}_L$ through one-qubit jumping (i.e., removal
of one excitation by the damping) are spanned by orthogonal
subspaces. Every one of these subspaces are given by two vectors
$\{\ket{R_{i0}},\ket{R_{i1}}\}$, $i=0,\ldots,4$, where $i=0$ denotes
no jump, $i=1,\ldots,4$ means a jump of (only) the first physical
qubit and so on. In total 10 vectors spans the no-jump and one-qubit
jump space. For no jump we have \beq
\ket{R_{00}}=\ket{0}_L,\quad\ket{R_{01}}=\ket{1}_L,\nonumber \eeq
and for the one-qubit jump it is easy to obtain that \bea
\ket{R_{10}}=\ket{0111},~~\ket{R_{11}}=\ket{0100}\nonumber,\\
\ket{R_{20}}=\ket{1011},~~\ket{R_{21}}=\ket{1000}\nonumber,\\
\ket{R_{30}}=\ket{1101},~~\ket{R_{31}}=\ket{0001}\nonumber,\\
\ket{R_{40}}=\ket{1110},~~\ket{R_{41}}=\ket{0010}\nonumber.\eea

In order to have a full four-qubit basis we can add the vectors:
\bea
\ket{R_{50}}=\left(\ket{0000}-\ket{1111}\right)/\sqrt{2},\nonumber\\\ket{R_{51}}=\left(\ket{0011}-\ket{1100}\right)/\sqrt{2}\nonumber,\\
\ket{R_{60}}=\left(\ket{1010}+\ket{0101}\right)/\sqrt{2},\nonumber\\\ket{R_{61}}=\left(\ket{1001}+\ket{0110}\right)/\sqrt{2}\nonumber,\\
\ket{R_{70}}=\left(\ket{1010}-\ket{0101}\right)/\sqrt{2},\nonumber\\\ket{R_{71}}=\left(\ket{1001}-\ket{0110}\right)/\sqrt{2}\nonumber.\eea
It is not important how one chooses these additional basis elements
since they do not correspond to any correctable error.

In this basis, let us introduce the following unitary transformation
to make the syndrome measurement more transparent:
\beq\hat{U}=\sum_{k=0}^{7}\sum_{i=0}^1\ket{i\textrm{Bin}(k)}\bra{R_{ki}},\label{syndrome}\eeq
where $\textrm{Bin}(k)$ is the three-bit, binary representation of
$k=0\ldots,7$. In an experimental realization, this transformation
can be incorporated in the syndrome measurement.

With the transformation given above one can approximately detect no
jump in any physical qubit, and one-qubit jumps by measuring the
three rightmost qubits in the computational basis. If the result is
000 there has been no jump, while the results 001, 010, 011, and 100
indicate that there have been a jump in qubits 1-4, respectively.
Actually, e.g., the result 010 can also indicate jumps in all of
qubits 2, 3, and 4 from the state $\ket{0}_L$. However, this code is
not designed to cope with multiple-qubit jumps, so it will act on
this syndrome as if it was caused by a single qubit jump in qubit 2.

Once the syndrome is measured we will apply the corresponding
recovery operation (summarized in the table below): If we measure
$\textrm{Bin}(k)$ for $k=0,\ldots,4$, we will preserve the possible
superposition carried by the first qubit by implementing the
transformation
$\hat{R}_k=\sum_{i=0}^1\ket{i}_L\bra{i\textrm{Bin}(k)}$. If we
happen to measure the syndromes $\textrm{Bin}(k)$ for $k=5,6,$ or
$7$, we know multiple (but not which) jumps have occurred. Hence we
cannot correct the state but chose to project the system to the
state $\hat{I}_L/2$, where
$\hat{I}_L=\ket{0}_L\bra{0}_L+\ket{1}_L\bra{1}_L$ is the identity
operator in the codeword space. After correction, the state is
always in a four-dimensional codeword-space enabling one to compute
(or measure) the fidelity to the initial state or the concurrence in
the standard fashion.
\begin{table}[h]
\caption{The syndromes and corresponding recovery transformations.}
\begin{center}
\begin{tabular}{|c|c|}
\hline Syndrome&Recovery transformation\\
\hline
$\textrm{Bin}(0)=000$&${\R}_0=\sum_{i=0}^1\ket{i}_L\bra{i\textrm{Bin}(0)}$\\
$\textrm{Bin}(1)=001$&${\R}_1=\sum_{i=0}^1\ket{i}_L\bra{i\textrm{Bin}(1)}$\\
$\textrm{Bin}(2)=010$&${\R}_2=\sum_{i=0}^1\ket{i}_L\bra{i\textrm{Bin}(2)}$\\
$\textrm{Bin}(3)=011$&${\R}_3=\sum_{i=0}^1\ket{i}_L\bra{i\textrm{Bin}(3)}$\\
$\textrm{Bin}(4)=100$&${\R}_4=\sum_{i=0}^1\ket{i}_L\bra{i\textrm{Bin}(4)}$\\
\cline{2-2}
$\textrm{Bin}(5)=101$& Project onto\\
$\textrm{Bin}(6)=110$&$\hat{I}_L/2$ \\
$\textrm{Bin}(7)=111$&\\\hline
\end{tabular}
\end{center}
\end{table}

\section{Does quantum error correction prevent entanglement sudden death?}

As pointed out in the introduction, it has been shown that ESD is
present in different scenarios for some bipartite states
\cite{diosi,dodd,yu1,esd,esd1,xstates,esd2,eberly,almeida}. In
particularl, the so-called ``X-states'' \cite{esd1} have been widely
studied \cite{xstates}. They are called X-states because their
composite density matrix has non-zero elements only in its diagonal
and antidiagonal, if expressed in the basis
$\{\ket{00},\ket{01},\ket{10},\ket{11}\}$. They also have the
property that the corresponding density matrix preserves the X form
when evolving under the action of certain system dynamics, and in
particular under dissipation. The class contain the Bell-states and
the Werner-states and therefore merits studying. The class also
encompasses both states that succumb to ESD and states that does
not. To limit our study somewhat, we will work with the Bell-like
X-states of the form \bea\label{phi}
\ket{\phi_0}&=&\cos\al\ket{11}+e^{i\beta}\sin\al\ket{00},\\
\label{psi}
\ket{\psi_0}&=&\cos\al\ket{10}+e^{i\beta}\sin\al\ket{01}.\eea For
comparison, we shall also study the separable, non-X-state
\beq\ket{\varphi_0}~=~\cos\alpha\ket{11}+e^{i\beta}\sin\alpha\ket{10}.\label{unent}\eeq
The state $\ket{\phi_0}$ will suffer from ESD when
$\vert\tan\alpha\vert<1$. This was experimentally demonstrated in
\cite{almeida}. Meanwhile, $\ket{\psi_0}$ does not, and
$\ket{\varphi_0}$ certainly does not, as is separable.

We shall now try to protect the information and entanglement carried
by the states (\ref{phi})-(\ref{unent}) from the action of
independent amplitude damping channels, and particularly from ESD
(if present) by encoding each qubit using the codewords
(\ref{ceroL}) and (\ref{unoL}), so that we obtain the logical states
\bea\label{phiL}
\ket{\phi}&=&\cos\al\ket{11}_L+e^{i\beta}\sin\al\ket{00}_L,\\
\label{psiL}
\ket{\psi}&=&\cos\al\ket{10}_L+e^{i\beta}\sin\al\ket{01}_L,\\
\ket{\varphi}&=&\cos\alpha\ket{11}_L+e^{i\beta}\sin\alpha\ket{10}_L.\label{unentL}\eea
To characterize to what extent the states are ``protected'' by
coding and error correction we will use the fidelity to the initial
state and the concurrence.

After damping and error correction the fidelities between the states
(\ref{phiL})-(\ref{unentL}) and the corresponding undamped states,
are given (to the second order in $\gamma$) by: \bea
{\F}_{\phi}&=&1-\frac{\gamma^2}{2}\left(7-3\cos2\al-\cos4\al\right)+O(\gamma^3),\label{fidphi}\\
{\F}_{\psi}&=&1-\frac{\gamma^2}{2}\left(7-\cos4\al\right)+O(\gamma^3),\label{fidpsi}\\
{\F}_{\varphi}&=&1-\frac{\gamma^2}{8}\left(17-6\cos2\al+\cos4\al\right.\nonumber\\&&\left.-2\cos2\beta\sin^22\al\right)+O(\gamma^3).\label{fidunen}\eea
Meanwhile, the fidelities between the initial and damped
corresponding uncoded states are: \bea
{\F}_{\phi_0}&=&1-2\gamma\cos^2\al+\gamma^2\cos^2\al,\label{fidphi0}\\
{\F}_{\psi_0}&=&1-\gamma\label{fidpsi0},\\
{\F}_{\varphi_0}&=&1-\frac{\gamma}{8}(11+4\cos2\al+\cos4\al)\nonumber\\
&&+\frac{\gamma^2}{8}\cos^2\al(3+5\cos2\al)+O(\gamma^3).\label{fidunen0}\eea

\begin{center}
\begin{figure}
\includegraphics[width=0.45\textwidth]{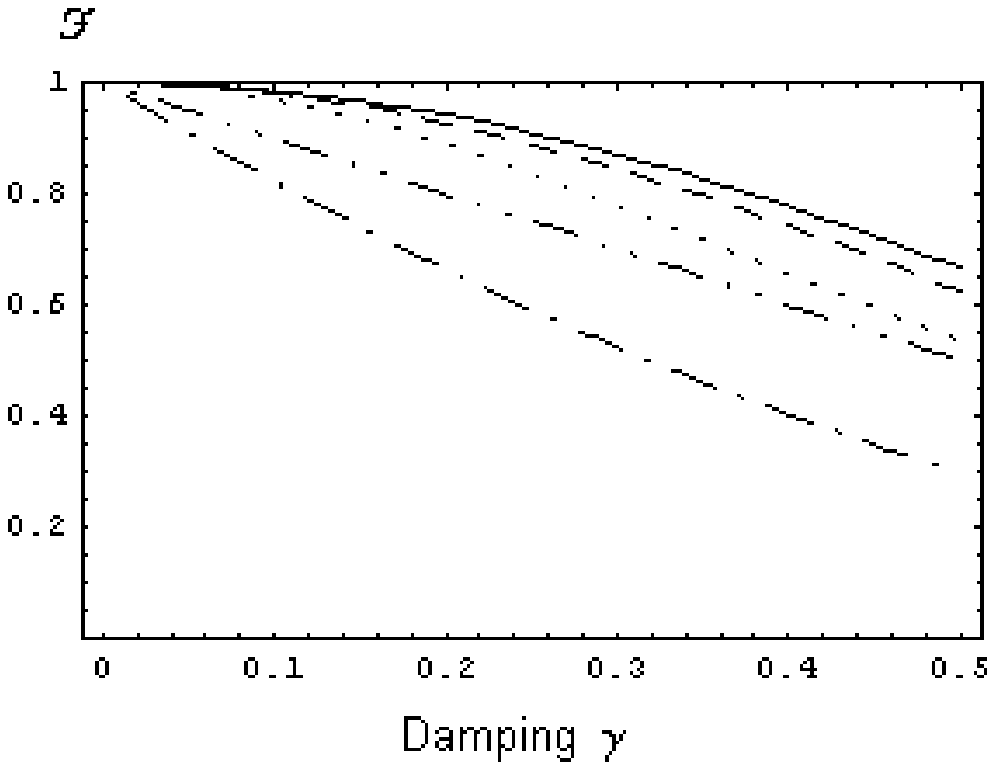}(a)
\includegraphics[width=0.45\textwidth]{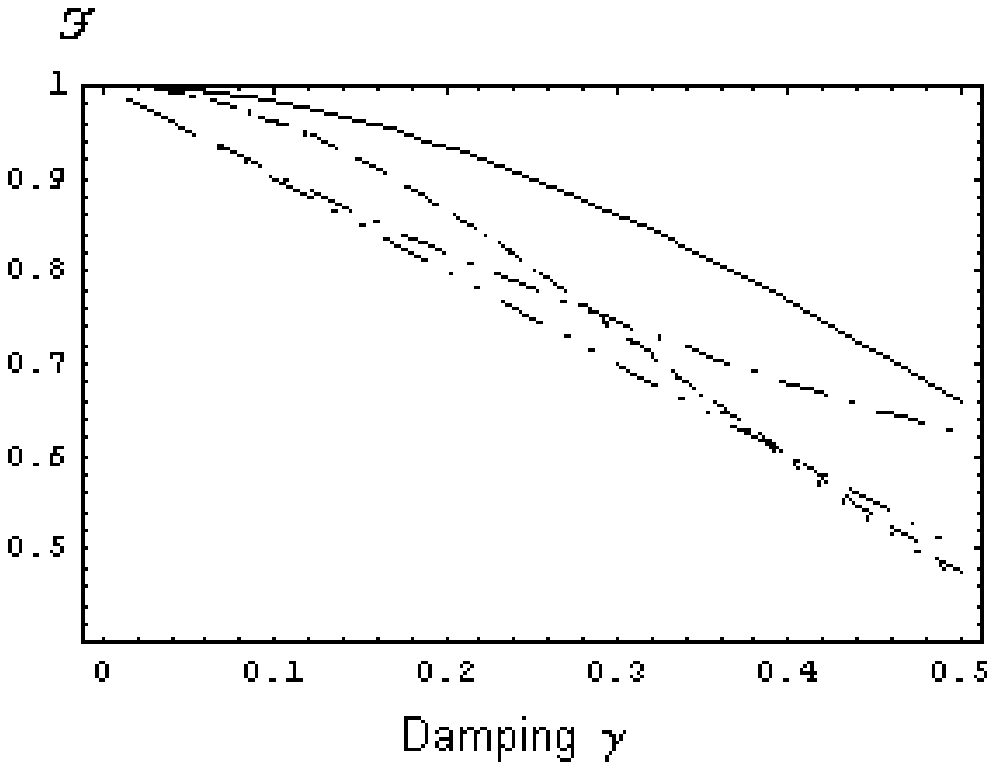}(b)
\caption{Plot of the fidelity after recovering for the states
$\ket{\phi}$ (dashed), $\ket{\psi}$ (dotted), and $\ket{\varphi}$
(continuous) together with the fidelity for the uncoded states
$\ket{\phi_0}$ (dash-dotted) and $\ket{\psi_0}$ (dash-double
dotted). The parameters are $\beta=0$, (a) $\al=\pi/12$ and (b)
$\al=\pi/4$. (In (b) the curves for $\ket{\phi}$ and $\ket{\psi}$
almost coincide.)}\label{fig1}
\end{figure}
\end{center}
As expected, the loss of fidelity obtained is of the order of
$\gamma^2$ for the error corrected states while it is linear for the
uncoded states. The exception is ${\F}_{\phi_0}$ that approaches
unity when $\al\rightarrow\pi/2$. This is to be expected since, in
this limit, the state is the trivial, nonevolving, ground state.

The expressions for the concurrence ${\cal C}$ of the states are
analytically ``messy'', and therefore we have chosen to simply plot
the exact, numerically results. The plotted fidelities are also
based on the exact expressions and not on the series expansions
(\ref{fidphi})-(\ref{fidunen0}). Since the dependence of on $\beta$
is weak, we have, rather arbitrarily, set $\beta=0$ when making the
plots. In Fig. \ref{fig1} we can compare the fidelities
(\ref{fidphi})-(\ref{fidpsi0}). In general, the fidelity as a
function of the damping will be larger for the unentangled state
$\ket{\varphi}$ than for the states $\ket{\phi}$ and $\ket{\psi}$.
The fidelity for the state $\ket{\phi}$ is larger than the one for
$\ket{\psi}$ for $\al<\pi/4$, for $\al<\pi/4$ the opposite happen,
being roughly the same when $\al=\pi/4$.

\begin{center}
\begin{figure}
\includegraphics[width=0.45\textwidth]{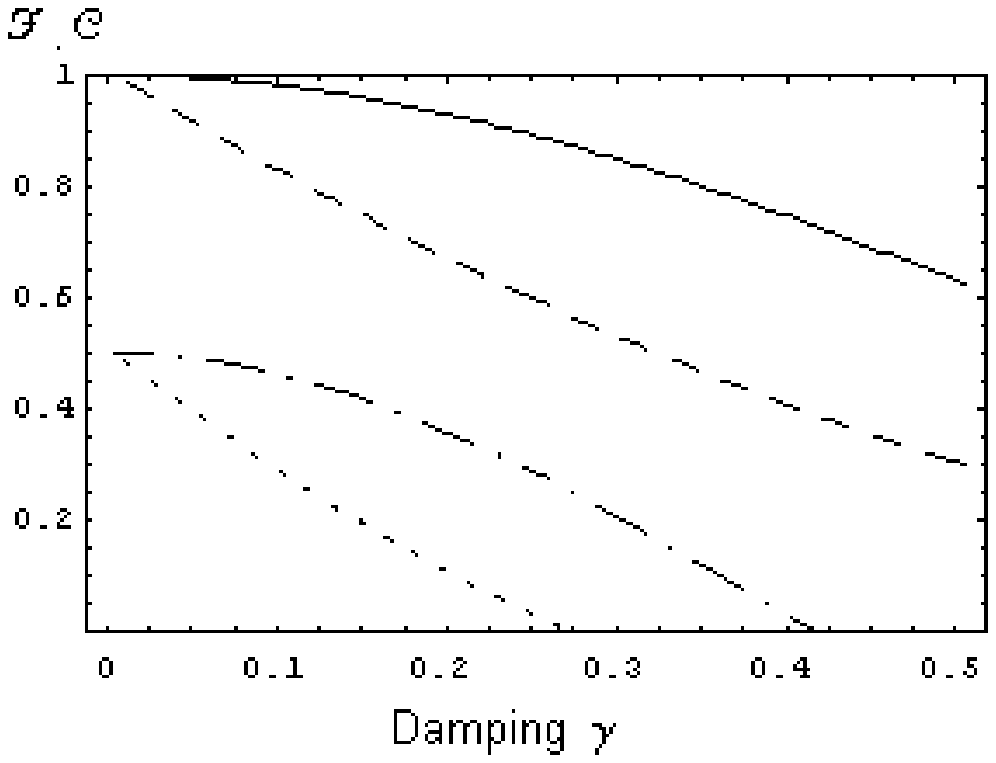}(a)
\includegraphics[width=0.45\textwidth]{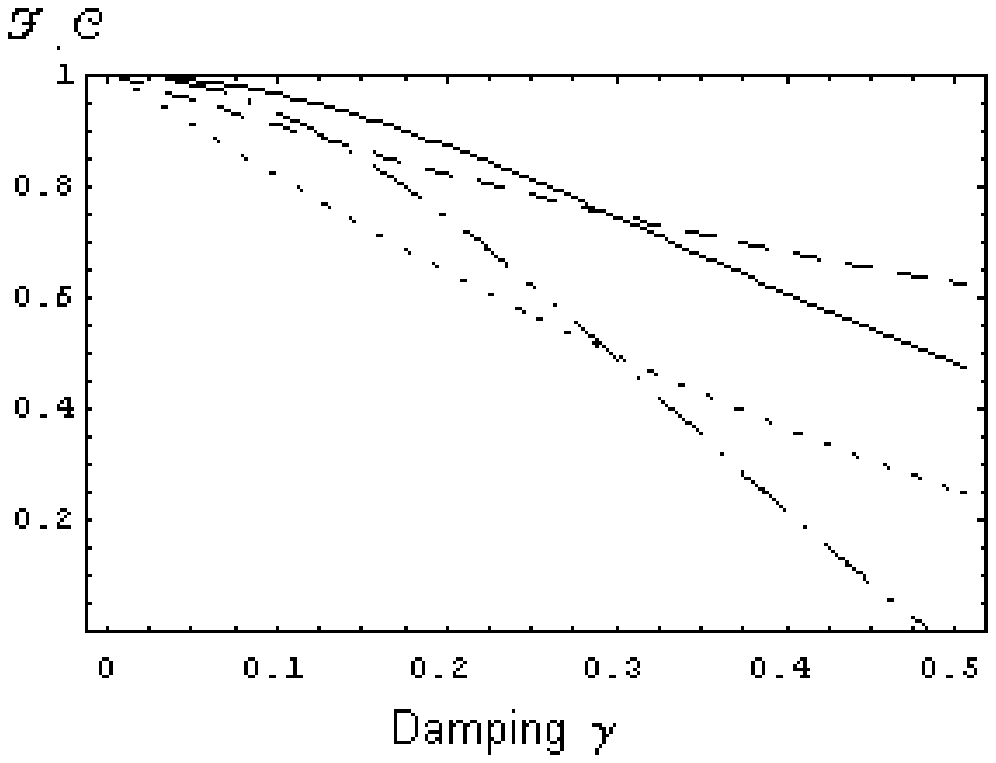}(b)
\caption{Plot of the fidelity (continuous) and the concurrence
(dash-dotted) for the state $\ket{\phi}$ after amplitude damping and
recovery, together with the fidelity (dashed) and the concurrence
(dotted) for the uncoded state $\ket{\phi_0}$. The parameters are
$\beta=0$, (a) $\al=\pi/12$ and (b) $\al=\pi/4$.}\label{fig2}
\end{figure}
\end{center}

In Fig. \ref{fig2} we plot the fidelities and concurrences for the
coded state $\ket{\phi}$ after error correction, and the
corresponding uncoded state $\ket{\phi_0}$ as a function of the
damping. In Figure \ref{fig2} (a) it is seen that the quantum error
correction makes the state almost twice as tolerant to damping
before the entanglement vanishes. However, this is not a general
truth, and this can be seen in \ref{fig2} (b) where the coded state
suffer from ESD while the uncoded state does not (because for the
uncoded state ${\cal C} = 0$ only as $\gamma = 1$). What is also
clear is that for small values of $\gamma$, quantum error correction
increases both the fidelity and the entanglement.

\begin{center}
\begin{figure}
\includegraphics[width=0.45\textwidth]{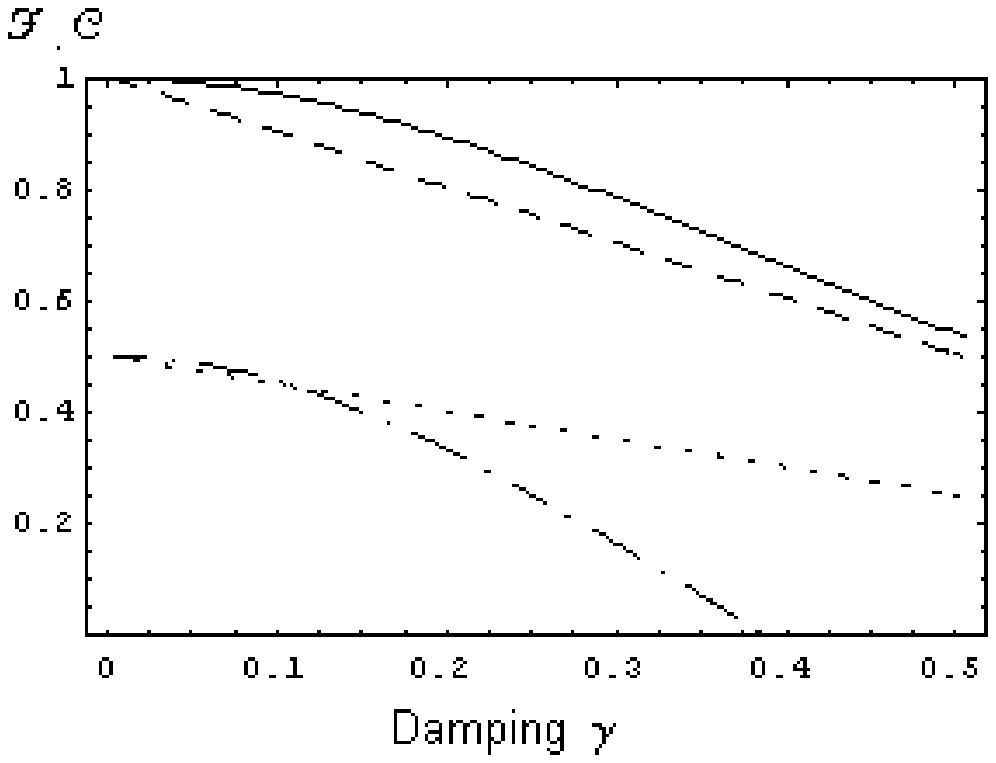}(a)
\includegraphics[width=0.45\textwidth]{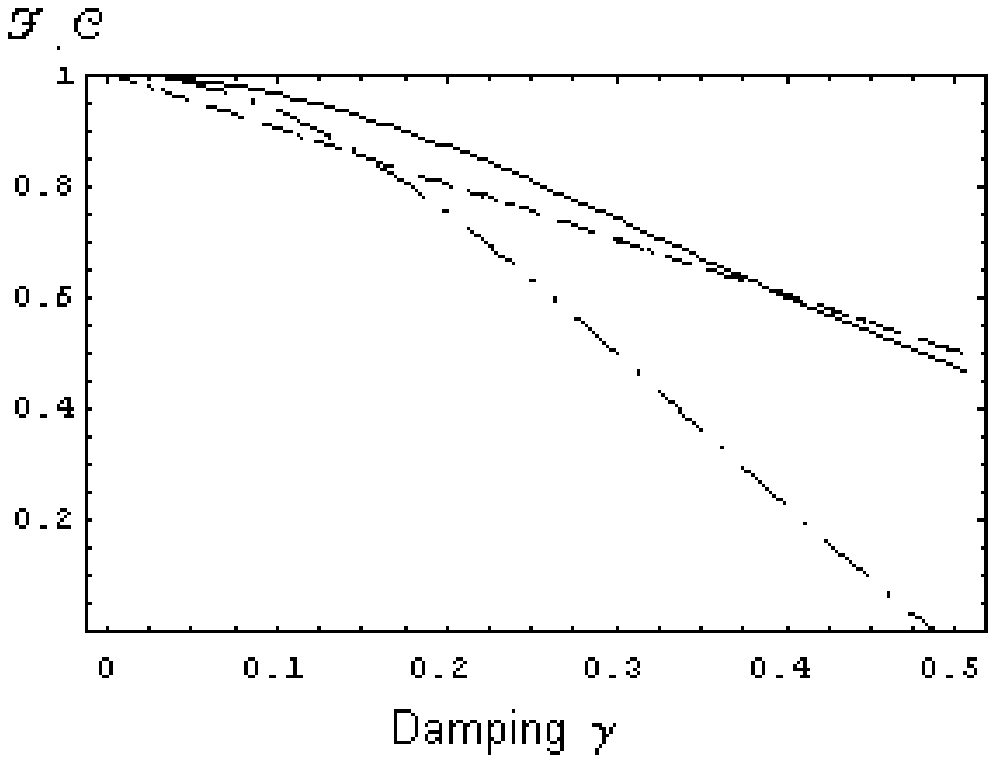}(b)
\caption{Plot of the fidelity (continuous) and the concurrence
(dash-dotted) for the state $\ket{\psi}$ after amplitude damping and
recovery, together with the fidelity (dashed) and the concurrence
(dotted) for the uncoded state $\ket{\psi_0}$. The parameters are
$\beta=0$, (a) $\al=\pi/12$ and (b) $\al=\pi/4$.}\label{fig3}
\end{figure}
\end{center}

In Fig. \ref{fig3} we plot the fidelities and concurrences for the
logical state $\ket{\psi}$ and for the corresponding uncoded state
$\ket{\psi_0}$. When $\al=\pi/4$, the concurrence and the fidelity
for the uncoded state $\ket{\psi_0}$ have the same value,
$1-\gamma$, and the curves in Figure \ref{fig3}(b), are not
distinguishable. Here again, we see that for large values of
$\gamma$ the coded state loses its entanglement in a finite time,
while the uncoded state only looses its entanglement asymptotically
so that ${\cal C} = 0$ only for $\gamma = 1$. However, for small
values of $\gamma$ both the fidelity and the concurrence is
increased by the use of quantum error correction, as expected.

\section{Conclusions}

With a few illustrative examples we have studied the possibility of
``protecting'' two-qubit, pure, Bell-like states from the action of
amplitude damping. As we have seen, quantum error correction
succeeds in protecting the state to first order in the dissipation
parameter, but when the dissipation increases beyond a certain
point, quantum error correction often makes matters worse and not
better. While longer and more complex codes may protect the states
to higher orders in the dissipation parameter, we conjecture that
the change is quantitative rather than qualitative compared to
simpler codes. In particular, we find that quantum error correction
will not protect a state from entanglement sudden death. On the
contrary, states that do not succumb to entanglement sudden death
may do so after they are coded.

Our examples also show what is perhaps obvious, but not much
discussed, namely that fidelity is insufficient to quantify quantum
processes and protocols. In all cases, the fidelity decreases
asymptotically with increasing imperfections. However, the resource
that is often of primary interest in quantum information, the
entanglement, does not decrease asymptotically. After suffering a
finite amount of dissipation the entanglement is gone, while the
fidelity may still seem comfortably high. The fidelity may thus be
deceptive as a figure of merit.

While we have only presented results for the $[4,1]$-code (applied
to the two qubits separately), we have also made similar analyses
for the $[5,1]$-code given in \cite{paz,5qubit} and the $[6,2]$-code
in \cite{Fletcher} (that encodes both qubits together). These codes
will give marginally better results in the fidelity and concurrence,
with similar syndrome measurement and recovery operations, but the
qualitative behavior, discussed above, remains. However, the
$[5,1]$-code is a more complex code and will be more difficult to
implement in practice. The $[6,2]$ code, in turn, is a nonlocal code
and employs non-local recovery operations. Hence, its implementation
requires creation of entanglement over the same physical distances
as the entangled qubits. This is a substantial drawback of the code.

Our final conclusion is not surprising. In order to combat loss of
entanglement due to dissipation one must employ quantum error
correction. It is important, however, to apply the recovery
operation while the dissipation is still minor, say, $\gamma \leq
0.1$. For values of $\gamma \geq 0.5$ the entanglement may already
have vanished, and will typically have done so if coding for quantum
error correction has been used.

\section*{Acknowledgements}
This work was supported by the Swedish Foundation for International
Cooperation in Research and Higher Education (STINT), the Swedish
Research Council (VR), and the Swedish Foundation for Strategic
Research (SSF).

\end{document}